\documentclass[letterpaper,twocolumn]{article}

\usepackage[T1]{fontenc}

\usepackage{geometry}
\usepackage{setspace}

\usepackage[style = chem-acs]{biblatex}
\usepackage{graphicx}

\usepackage{amsmath}
\usepackage{amssymb}
\usepackage{bm}

\usepackage{float}
\newfloat{scheme}{htbp}{los}
\floatname{scheme}{Scheme}
\floatname{chart}{Chart}
\newfloat{graph}{htbp}{loh}

\usepackage{chemformula} 
\usepackage[version = 4]{mhchem} 

\usepackage{authblk}
\author[1]{Siyu Duan}
\affil[1]{Department of Physics, TU Dortmund University, 44227 Dortmund, Germany}
\author[2]{Lili Shi}
\affil[2]{NEST, CNR-Istituto Nanoscienze and Scuola Normale Superiore, Piazza San Silvestro 12, Pisa 56127, Italy}
\author[1]{Patrick Pilch}
\author[1]{Anneke Reinold}
\author[1]{Sergey Kovalev}
\author[3,*]{Renato~M.~A.~Dantas}
\affil[3]{Department of Physics, University of Basel, 4056 Basel, Switzerland}
\author[4,5,6]{Yunkun Yang}
\affil[4]{State Key Laboratory of Surface Physics and Department of Physics, Fudan University, Shanghai, China}
\affil[5]{Shanghai Research Center for Quantum Sciences, Shanghai, China}
\affil[6]{Beijing Academy of Quantum Information Sciences, Beijing, China}
\author[4,5,*]{Faxian Xiu}
\author[2,*]{Miriam Serena Vitiello}
\author[1,*]{Zhe~Wang}

\title{Giant enhancement of terahertz high-harmonic generation by cavity engineering of a three-dimensional Dirac semimetal}
\date{*Email: renatomiguel.alvesdantas@unibas.ch\\
               faxian@fudan.edu.cn\\
                 miriam.vitiello@sns.it\\
                 zhe.wang@tu-dortmund.de
}

\begin{document}

\twocolumn[
\maketitle

\begin{center}
\begin{minipage}{0.9\textwidth}

\begin{abstract}
  We report on time-resolved ultrafast terahertz high-harmonic generation of strong field driven-dynamics of many-body Dirac fermions.
We demonstrate an experimental realization of near saturation regime of the high-order nonlinear responses with a giant enhancement of terahertz third- and fifth-order harmonic yields by cavity-engineering a three-dimensional Dirac semimetal Cd$_3$As$_2$.
By fabricating a designed structure of metasurface microcavities on a nanometer Cd$_3$As$_2$ thin film, we significantly enhance the near-field intensity of a picosecond terahertz excitation pulse in resonance with the microcavity eigenmode.
The strong terahertz field drives the far-from-equilibrium Dirac fermions deeply into a nonperturbative regime, leading to the observation of near-saturation high-harmonic emission.
Our experimental results confirm the predictions by Boltzmann transport theory, and substantiate a field-driven kinetic description of the strong nonthermal
nonlinearity at the terahertz frequencies.
\end{abstract}

\vspace{0.5em}

\end{minipage}
\end{center}

\vspace{1cm}
]

Matter driven far from thermal equilibrium may exhibit properties on transient time scales that are very different from the responses of its thermal equilibrium state \cite{Torre2021}.
A prominent example is the high-order harmonic generation (HHG) of atomic or molecular gases. Under the drive of a strong laser electric field in the near-infrared range, typically of 10--100~MV/cm, the atoms or molecules emit very short pulses with a duration down to attoseconds, laying the foundation for the studies of attosecond electron dynamics \cite{Corkum2007,Goulielmakis2022,Heide2024}.
A three-step scenario including tunneling, acceleration, and recombination has been well established to describe the dynamical processes of atomic and molecular HHG.
Since the typical band gaps of a solid-state insulator is of a few electronvolts (eV), comparable to the atomic energy scale, near-infrared laser-excited HHG has been observed also in insulators with the corresponding dynamics similarly described by three main processes involving interband excitation, dynamical Bloch oscillations, and atomic-like recollision \cite{Heide2024}.

In this work, we investigate HHG in a distinct paradigm: strong terahertz (THz) field-driven high-harmonic generation of Dirac fermions.
Compared with the infrared laser induced HHG, the fundamental difference arises from the facts that 
(i) the THz photon energy is a few milli-electronvolts (meV), which is much smaller than the typical band gap of an insulator or semiconductor, and 
(ii) the available THz electric field strength of  0.1--1~MV/cm is about two orders of magnitude weaker than the infrared laser field.
For these reasons, in a Dirac semimetal we can Pauli-block THz interband excitation by shifting the Fermi level with charge doping, thereby focusing on the intraband dynamics of the Dirac fermions driven by strong terahertz field far from thermal equilibrium.

For two-dimensional (2D) Dirac systems such as graphene and topological insulators (e.g. Bi$_2$Se$_3$) \cite{Hafez2018,Deinert2021,Tielrooij2022}, THz pulse driven HHG has been ascribed to ultrafast heating of the Dirac fermions \cite{Hafez2018,Deinert2021,Tielrooij2022}. By fabricating metallic gratings on graphene, the nonlinear susceptibility corresponding to third-harmonic generation (THG) was enhanced by one order of magnitude at low incidence fields \cite{Deinert2021}, whereas at high-incidence fields the system quickly entered the saturation regime due to rapid heat accumulation where the efficiency of harmonic generation was even lower than in a bare graphene \cite{Tielrooij2022}. For Bi$_2$Se$_3$ the enhancement of THG due to fabricated metallic gratings was also observed at low THz fields but weaker than in graphene, because the bulk states in Bi$_2$Se$_3$ provide an additional channel to relax the THz heating effects on the surface states \cite{Tielrooij2022}. 
However, at high fields the enhancement of THG in Bi$_2$Se$_3$ due to the metallic gratings is negligibly small, although the THG is still far from being saturated \cite{Tielrooij2022}.

A different scenario is studied in this work on the following aspects: 
Firstly, we investigate THz field driven kinetics rather than assuming a thermodynamic scenario of heating effects.
Secondly, we study a three-dimensional Dirac semimetal beyond two dimensions. 
Thirdly, instead of a metallic grating, we fabricate a resonator cavity metasurface that can resonantly enhance THz field at the Dirac material.
We find that a giant enhancement of THz HHG in the 3D Dirac semimetal is achieved by using the resonator cavity metasurface, which can be described by nonlinear field-driven kinetics of the Dirac fermions.

\begin{figure}[t]
\centering
\includegraphics[width=\columnwidth]{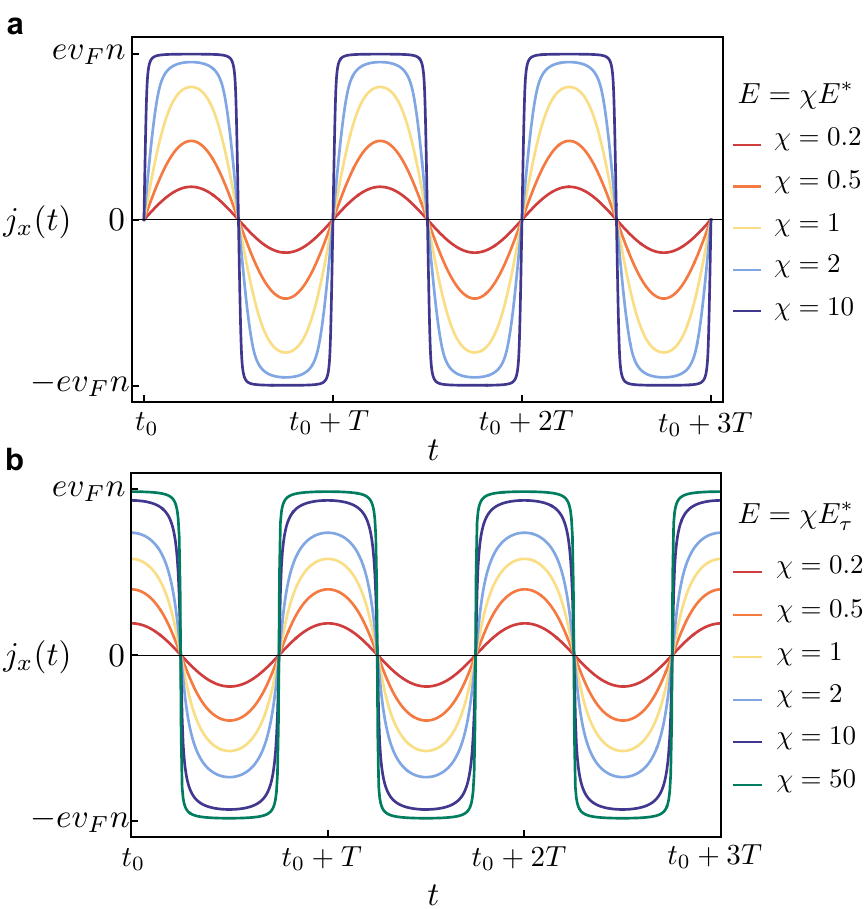}
\caption{\label{fig:Theory}
Time-dependent current density along the field direction, $j_x(t)$, for a single Weyl node driven by an electric field $\bm{E}(t)=E\cos(\omega t)\,\hat{\bm{x}}$ in (\textbf{a}) the collisionless limit ($\tau\rightarrow\infty$) [Eq.~(\ref{eq:collisionless})] and (\textbf{b}) the collision-dominated limit ($\omega\tau\ll 1$) [Eq.~(\ref{eq:w_collision})]. The field amplitude is parametrized as $E=\chi E^*$ in (\textbf{a}), with $\chi=0.2,\,0.5,\,1,\,2,\,10$ and $E^*=\frac{\mu\omega}{e v_F}$, and as $E=\chi E^*_{\tau}$ in (\textbf{b}), with $\chi=0.2,\,0.5,\,1,\,2,\,10,\,50$ and $E^*_{\tau}=\frac{\mu}{e v_F \tau}$. For the numerical evaluation, we use $f=1/T=\omega/(2\pi)=0.35~\mathrm{THz}$, $\mu=118~\mathrm{meV}$, $v_F=7.8\times 10^{5}~\mathrm{m/s}$, and $\tau=10~\mathrm{fs}$. Here $t_0$ denotes an arbitrary reference time.
}
\end{figure}

The far-from-equilibrium response in Dirac semimetals induced by an intense THz laser field can be captured by the semiclassical Boltzmann transport theory (see e.g. ~\cite{Kovalev2020,Cheng20,Lim20,Dantas2021}). 
In this framework, the state of a Dirac semimetal far from equilibrium is encoded in the distribution function $f (t, \bm{p})$, whose time evolution is governed by the Boltzmann equation
\begin{equation}
\partial_t f(t,\bm{p}) = e \bm{E}(t) \cdot \nabla_{\bm{p}} f(t, \bm{p}) + \frac{f_0 (\bm{p}) - f(t,\bm{p})}{\tau}. 
\end{equation}
The first term on the right-hand side describes the dynamics imposed on the massless quasiparticles by a homogeneous THz electric field $\bm{E}(t)$, with $e$ denoting the elementary charge. The second term captures relaxation processes due to scattering events, which are modeled here within the relaxation-time approximation by a single phenomenological parameter $\tau$. These relaxation processes drive the system back toward thermal equilibrium, where the distribution is given by the Fermi-Dirac function $f_0(\bm{p})$.

To establish the fundamental response of a Dirac semimetal to a THz driving field, we focus on the response current of a single Weyl node and consider first the collisionless limit $\tau \rightarrow \infty$.
In this limit, the relaxation term in the Boltzmann equation vanishes and the solution reads $f(t,\bm{p})= f_0(\bm{p}-e \bm{\Delta}(t,0))$, describing a many-body system whose quasiparticles are accelerated by the external field and acquire a momentum shift $e \bm{\Delta}(t,0)= - e \int^t_0 \bm{E}(s) ds$.
At zero temperature, the time-dependent current density $\bm{j}(t)$ can be evaluated analytically \cite{Dantas2021}
\begin{equation}\label{eq:collisionless}
\bm{j}(t) =- e \, v_F n \, \bm{\mathcal{F}} \left(\frac{v_F e \bm{\Delta}(t,0)}{\mu} \right),
\end{equation}
where $n= \tfrac{\mu^3}{6 \pi^2 \hbar^3 v_F^3}$ is the carrier density, $\mu$ is the chemical potential, $v_F$ is the Fermi velocity, and
\[
\bm{\mathcal{F}} \left(\bm{x}\right) =
\begin{cases}
(1 -\tfrac{1}{5} x^2) \, \bm{x}, &  | \bm{x} | \le 1, \\[4pt]
(\tfrac{1}{x} - \tfrac{1}{5} \tfrac{1}{x^3}) \, \bm{x}, & | \bm{x} | > 1.
\end{cases}
\] 
These results establish two different regimes of the nonlinear response.
In the weak-field regime where the field-induced momentum shift is smaller than the Fermi momentum, i.e. $e | \bm{\Delta}(t,0)| \le \frac{\mu}{v_F}$, the nonlinear response is perturbative and the current density $\bm{j}(t)$ contains only linear and cubic terms in $\bm{\Delta}(t,0)$ producing only the fundamental and third harmonics \cite{Dantas2021,Lim20}. 
In contrast, when the momentum shift exceeds the Fermi momentum at strong fields, i.e. $e | \bm{\Delta}(t,0)| > \frac{\mu}{v_F}$, the nonlinear response becomes nonperturbative, leading to the radiation of all odd-order harmonics \cite{Dantas2021,Lim20}.
Moreover, in the strong-field limit ($x\gg 1$), $| \bm{\mathcal{F}}(\bm{x})|\to 1$ and $\bm{\mathcal{F}}(\bm{x})$ remains parallel to $\bm{\Delta}(t,0)$, which oscillates in time under the driving electric field, implying saturation of the response current and, consequently, of the high-harmonic generation even in the collisionless regime. 
For an ideal sinusoidal driving field $\bm{E}(t)=\bm{E}\cos(\omega t)$, the crossover from the perturbative to the nonperturbative regime occurs at the critical field $E^*=\frac{\mu\omega}{e v_F}$, and as shown in Figure~\ref{fig:Theory}a, the response current nearly saturates for field amplitudes $E\gtrsim 10\,E^*$ yielding a nearly rectangular waveform in the time domain.

When the relaxation time $\tau$ is not infinite but rather shorter than the THz cycle (i.e. $\omega\tau \ll 1$ for a THz drive frequency $\omega$), the relaxation term in the Boltzmann equation is essential and the zero-temperature response current can be written as \cite{Pilch2025}
\begin{align}\label{eq:w_collision}
\footnotesize
\bm{j} (t) \approx  \frac{-e \, v_F n }{1-e^{-T/\tau}} \int_{-T}^{0} \frac{du}{\tau}  \,e^{u/\tau} \bm{\mathcal{F}} \left(\frac{v_F e \bm{\Delta}(t,t+u)}{\mu} \right)
\end{align} 
The inclusion of relaxation processes strongly modifies the form of the out-of-equilibrium distribution, resulting in an overall reduction in the efficiency of the response. 
Considering again the driving field $\bm{E}(t)=\bm{E}\cos(\omega t)$, the crossover from the perturbative to the nonperturbative regime occurs around $E^{*}_{\tau}\sim \frac{\mu}{e v_F \tau}$. 
Due to the presence of $\bm{\mathcal{F}}(\bm{x})$ in the integrand, the response current in the collision-dominated regime is also expected to saturate for sufficiently strong driving fields (Figure~\ref{fig:Theory}b). Consequently, the high-harmonic response is expected to eventually saturate.
Further theoretical analysis of temperature dependence and THz driving field dependence of the HHG yields qualitatively similar results, which is provided in the Supporting Information. According our previous simulation \cite{Kovalev2020}, under the drive of strong THz field, the value of the relaxation time in Cd$_3$As$_2$ is $\tau \approx 10$~fs, comparable with that in graphene \cite{Gierz2013}, hence the strong THz field-driven kinetics in Cd$_3$As$_2$ is in the collision-dominated regime.

\begin{figure*}[t]
\centering
\includegraphics[width=1\textwidth]{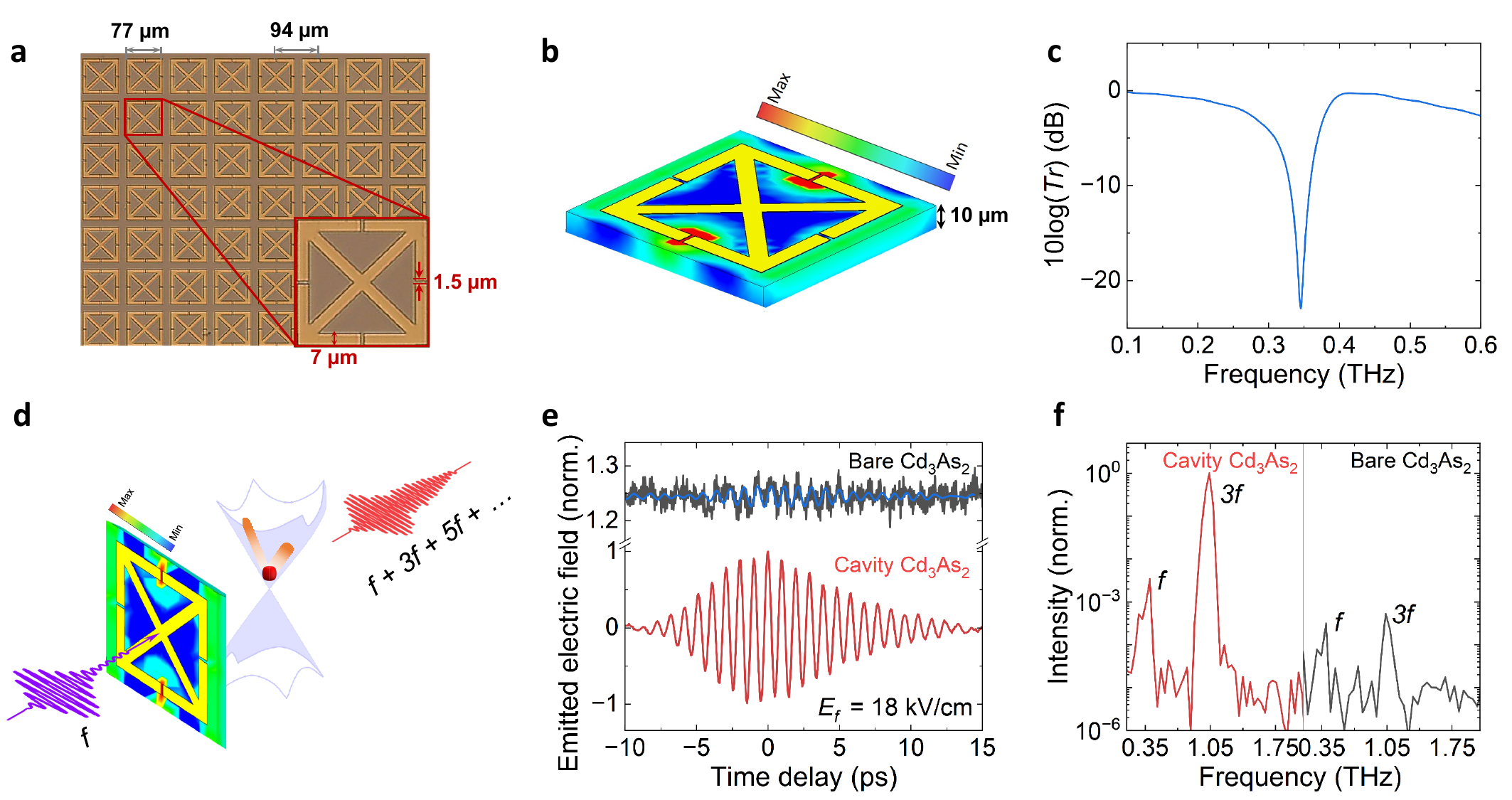}
\caption{\label{fig:Illustration}
(\textbf{a}) Image of regularly arranged gold split-ring resonators microcavities on top of Cd$_3$As$_2$ nanometer thin film.
(\textbf{b}) Simulated electric field strength in one cavity unit. A maximum enhancement of electric field by a factor of about 49 is achieved at the slit of the resonator.
(\textbf{c}) Simulated linear transmission spectrum of the resonator microcavity structure with a resonant frequency of 0.35~THz and a Q-factor of 54.
(\textbf{d}) Illustration of THz high-harmonic generation spectroscopic experiment of a cavity Dirac semimetal.
(\textbf{e}) Emitted electric field of bare Cd$_3$As$_2$ thin film sample (grey curve) and of the cavity sample (red curve) under the drive of an $f=0.35$~THz multicycle pulse with a peak driving THz field of $E_f=18$~kV/cm measured in far field.
The grey curve is shifted vertically for clarity.
(\textbf{f}) The corresponding frequency-domain spectra of the harmonic emission, exhibiting peaks at both the fundamental $f$ and the third-harmonic frequencies $3f=1.05$~THz. The data were recorded after a $3f$ bandpass filter.
}
\end{figure*}

To experimentally access the saturation regime of the extremely nonlinear THz response and investigate the strong-field driven nonthermal dynamics of Dirac fermions, we choose the three-dimensional Dirac semimetal Cd$_3$As$_2$.
Governed by massless quasiparticles of Dirac fermions, the electric transport properties in Cd$_3$As$_2$ feature an ultrahigh mobility in three dimensions \cite{Liang2015}, since Cd$_3$As$_2$ has no topologically trivial bands near the Fermi energy but only a pair of Dirac cones \cite{ZJWang13}.
Previous experimental studies have discovered THz high-harmonic generation in the nonperturbative regime of the THz electric field-driven nonlinear response \cite{Kovalev2020,Cheng20}.
The efficiency of the third-harmonic yield in Cd$_3$As$_2$ is greater than in graphene \cite{Cheng20} or Weyl semimetal \cite{Pilch2025}.
Nevertheless, the dependence of the observed third-harmonic radiation field $E_{3f}$ on the driving electric field $E_f$ follows $E_{3f} \propto E^\alpha_f$ with $\alpha = 2.1 - 2.6$ \cite{Kovalev2020,Cheng20}, which is not far above the perturbative response regime $E_{3f} \propto E^3_f$ that was found, for example, in the Weyl semimetal TaP \cite{Pilch2025}.

In order to drive the system deep into the nonthermal nonperturbative regime, we resonantly enhance the local THz electric field by fabricating a cavity metasurface structure on a nanometer-thick Cd$_3$As$_2$ thin film.
Constituted by periodically arranged gold split-ring cavities, the metasurface structure has an electromagnetic resonance frequency at $f = 0.35$~THz, corresponding to the peak power of our THz driving pulse.
As specified in Figure~\ref{fig:Illustration}a, 
the optimized parameters of the metasurface structure are derived by carrying out an electromagnetic simulation with a commercial software (CST Studio Suite 2022).
In the simulation, periodic boundary conditions were assumed for the in-plane \textit{x}- and \textit{y}-directions, while open boundaries for the out-of-plane \textit{z}-direction.
For our spectroscopic measurements, the high-quality Cd$_3$As$_2$ thin films with a thickness of about 60~nm were grown by molecular beam epitaxy \cite{Liu2015}.
The fabrication process of the cavity device started with spin-coating a layer of PMMA photoresist on the Cd$_3$As$_2$ thin film.
The array of the gold split-ring resonators was then patterned using electron beam lithography, which is followed by deposition of a 100-nm-thick gold film through thermal evaporation. 
The fabrication was completed through a lift-off process.
According to our simulation, the spacial distribution of electric field strength in one cavity unit is presented in Figure~\ref{fig:Illustration}b. 
A maximum enhancement of electric field by a factor of about 49 is achieved at the slit of the microresonators and the enhancement extends approximately 10~µm beneath the metasurface.
For our thin film of only 60~nm in thickness, the enhanced electric field penetrates the entire film. 
As shown in Figure~\ref{fig:Illustration}c, the simulated linear transmission spectrum of the metasurface structure has a sharp resonance at 0.35~THz with a Q-factor of 54.

\begin{figure*}[t]
\centering
\includegraphics[width=0.9\textwidth]{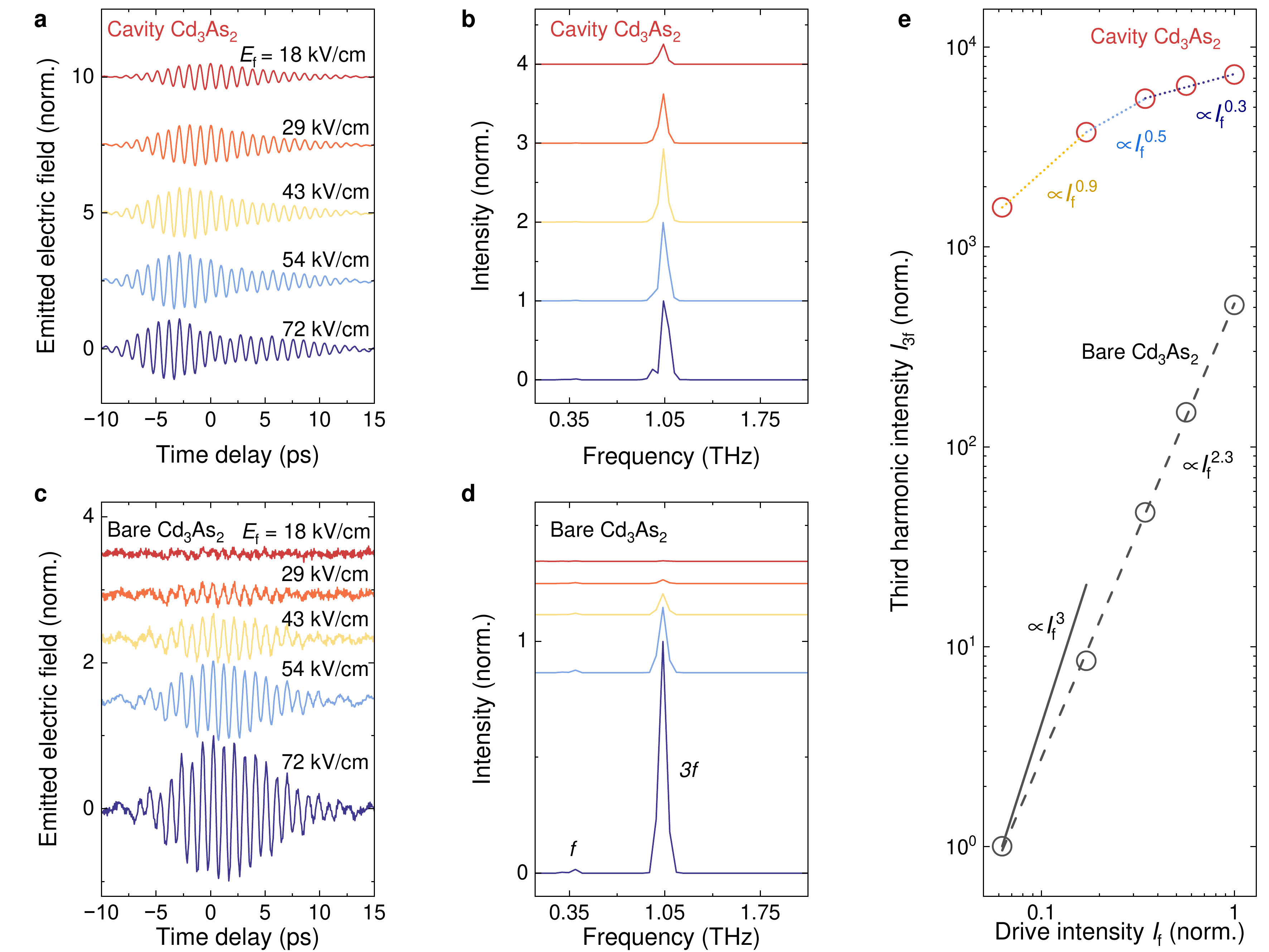}
\caption{\label{fig:THG}
(\textbf{a}) Time-resolved and (\textbf{b}) frequency-domain spectra of third-harmonic electric fields emitted from the cavity Cd$_3$As$_2$ sample for various peak driving electric field strengths $E_f$ measured in far field. 
(\textbf{c}) Time-resolved and (\textbf{d}) frequency-domain spectra of third-harmonic electric fields emitted from the bare Cd$_3$As$_2$ sample for various peak driving electric field strengths $E_f$ measured in far field.
The data are recorded after a $3f$-bandpass filter. 
The curves are shifted vertically for clarity.
(\textbf{e}) The intensity of the third-harmonic yield versus the driving pulse intensity for the cavity-engineered Cd$_3$As$_2$ sample and the bare Cd$_3$As$_2$ sample.
The bare Cd$_3$As$_2$ sample exhibits a nonperturbative dependence $I_{3f} \propto I^{2.3}_f$ (dashed line), in clear contrast to the perturbative behavior $I_{3f} \propto I^3_f$ (solid line). 
The dotted lines indicate power-law fits to the experimental data.
In the cavity-engineered sample, the third-harmonic yield is enhanced by more than three orders of magnitude at the lowest driving field, and approaches nearly saturation towards higher fields. 
}
\end{figure*}


The time-resolved ultrafast THz high-harmonic generation spectroscopic experiment performed in the present work is schematically illustrated in Figure~\ref{fig:Illustration}d.
Based on an 800~nm pulsed laser system (100~fs, 1~kHz), a multicycle $f=0.35$~THz driving pulse is prepared by using optical rectification in a LiNbO$_3$ single crystal \cite{Hebling02} and a $f$-bandpass filter \cite{Zhu2025}.
The THz driving pulse is linearly polarized and focused onto the Cd$_3$As$_2$-based samples. 
A bare Cd$_3$As$_2$ thin-film sample without metasurface was also measured as a reference.
The emitted THz radiation is measured in transmission configuration with the electric field recorded by electro-optic sampling in ZnTe crystal~\cite{Zhang95}.

The time-dependent emission of the bare Cd$_3$As$_2$ thin-film sample and of the microcavity-engineered sample is presented in Figure~\ref{fig:Illustration}e for a peak driving field of $E_f=18$~kV/cm measured in far field. The corresponding frequency-domain spectra are derived from Fourier transformation and displayed in Figure~\ref{fig:Illustration}f.
Although the THz driving field is rather weak (for comparison, a $\sim300$~kV/cm electric field was used in Ref.~\cite{Liu2012}), the bare Cd$_3$As$_2$ thin film emits clearly visible $3f$ radiation, which can be seen both in the time-domain and in the frequency-domain representations.
In Figure~\ref{fig:Illustration}e the blue curve indicates the $3f$ oscillation with the high-frequency noise numerically filtered out. 

In stark contrast, the microcavity-engineered sample exhibits a much more efficient third-harmonic yield.
Without any numerical filtering, the experimentally recorded time-domain data show a very strong $3f$ oscillation (see the red curve in Figure~\ref{fig:Illustration}e).  
In comparison with the bare Cd$_3$As$_2$ thin film, the enhancement of the third-harmonic generation due to the microcavities is more than three orders of magnitude (see Figure~\ref{fig:Illustration}f).
If we assume that at the lowest fluence the enhanced THG is confined exactly to the areas of the open slits of the split-ring resonators (see numerically simulated result in Figure~\ref{fig:Illustration}b),
the effective enhancement of the high-harmonic yields per unit area is even greater by another factor of about 420. 
Such a strong enhancement by more than five orders of magnitude per unit active area is the first important finding in the present work.
We note that the apparent stronger $3f$ emission than the $1f$ signal in Figure~\ref{fig:Illustration} is due to the suppression of the fundamental signal by a $3f$ bandpass filter in the detection scheme.


To systematically investigate the THz nonlinear responses of the cavity-engineered Dirac fermions, we measure the third-harmonic generation by varying the THz driving field strength.
The time-domain signals of the cavity-engineered sample and of the bare Cd$_3$As$_2$ thin film are shown in Figure~\ref{fig:THG}a and~\ref{fig:THG}c, respectively, while the corresponding frequency-domain spectra obtained by Fourier transformation are presented in Figure~\ref{fig:THG}b and~\ref{fig:THG}d.
For the bare Cd$_3$As$_2$ thin film, the third-harmonic yield increases very rapidly with increasing driving field strength (Figure~\ref{fig:THG}d).
While the cavity-engineered sample emits much stronger third-harmonic radiation, the field-dependent increase is less evident (Figure~\ref{fig:THG}b). These behaviours are clearly visible both in the time-domain and in the frequency-domain representations.
Moreover, in the time domain the cavity sample exhibits an evident shift of the maximum emission towards earlier time delays with a clear change of the shape of the THG pulse envelope at higher driving fields (Figure~\ref{fig:THG}a). 
In contrast, for the bare sample the shape of the THG pulse remains nearly the same (Figure~\ref{fig:THG}c).
Therefore, the change of the THG pulse shape is a characteristic of the strongly enhanced harmonic generation in the cavity sample.

The dependencies of the third-harmonic intensity $I_{3f}$ on the driving field intensity $I_f$ for the bare and the cavity-engineered samples are summarized in Figure~\ref{fig:THG}e for comparison.
The bare Cd$_3$As$_2$ thin film exhibits a clear power-law dependence, i.e. $I_{3f} \propto I^{2.3}_f$, as fitted by the dashed line in the log-log plot for the whole available range of THz intensity.
This confirms that the THz field driven nonlinear response of the Dirac fermions in Cd$_3$As$_2$ is in the nonperturbative regime, but still close to the perturbative response where $I_{3f} \propto I^{3}_f$ (see solid line in Figure~\ref{fig:THG}e).
The third-harmonic generation of the microcavity-engineered sample exhibits very different fluence dependent behaviours.
Firstly, in comparison with the bare Cd$_3$As$_2$ film, the enhancement of the third-harmonic intensity varies from being more than three orders of magnitude at the lowest fluence to merely more than one order of magnitude at the highest available fluence.
The corresponding power conversion efficiency $I_{3f}/I_f$ in the cavity sample changes from $8.3 \times 10^{-4}$ to $2.6 \times 10^{-4}$ with increasing fluence, whereas in the bare thin film from $5.3 \times 10^{-7}$ to $1.8 \times 10^{-5}$.
One may note that in comparison with graphene, even the bare Cd$_3$As$_2$ thin film can exhibit a higher conversion efficiency by more than one order of magnitude \cite{Cheng20}.
Secondly, the observed fluence dependence of the microcavity sample does not follow a single power law, but varies with increasing fluence.
Based on the available data, we can approximate the fluence dependent evolution by three power-law dependencies, i.e.  $I_{3f} \propto I^{0.9}_f$,  $I_{3f} \propto I^{0.5}_f$, and $I_{3f} \propto I^{0.3}_f$, as fitted by linear functions in the log-log plot and indicated by the dotted lines in Figure~\ref{fig:THG}e, a clearly continuous decrease of the power-law exponent towards higher fluences.
Thirdly, the nearly zero power-law dependence at the highest available THz field indicates that the cavity engineering enables us to enter deep into the extreme nonperturbative nonlinear regime - near saturation THz high-harmonic generation.
By using the THz metasurface microcavity we have reached this regime experimentally, corresponding to the upper physical limit of the nonlinear THz conversion efficiency \cite{Lim20,Dantas2021}.
These experimental findings confirm our simulation results based on the analysis of the Boltzmann transport theory (see Figure~\ref{fig:Theory} and the Supporting Information).

\begin{figure*}[t]
\centering
\includegraphics[width=0.87\textwidth]{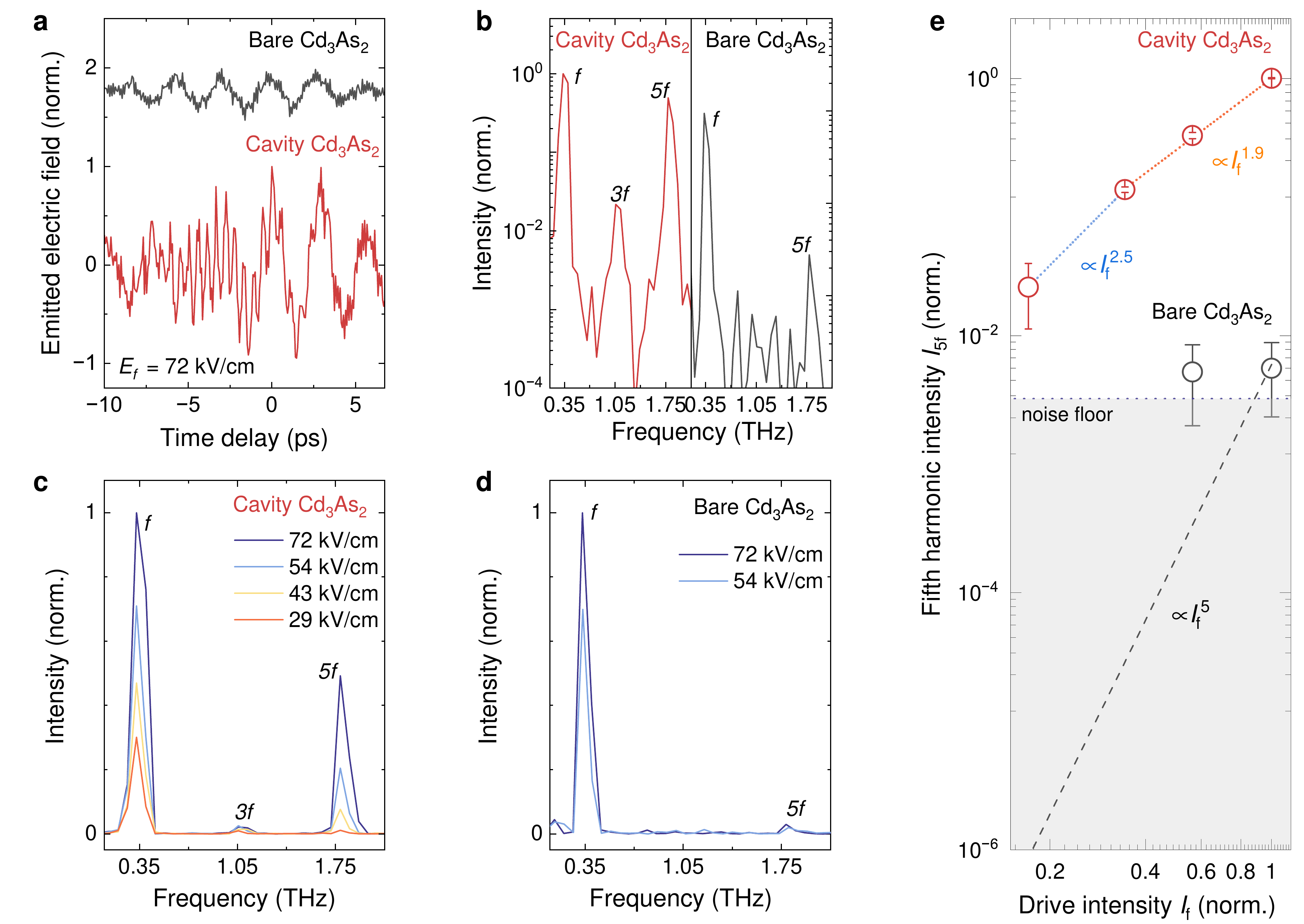}
\caption{\label{fig:FHG}
(\textbf{a}) Time-resolved emitted electric field and (\textbf{b}) the corresponding frequency-domain spectra from a THz field driven bare Cd$_3$As$_2$ sample and its microcavity heterostructure for a driving peak field of 72~kV/cm (measured in far field) at the frequency $f=0.35$~THz.
The emitted electric fields are recorded after a $5f$-bandpass filter. 
(\textbf{b}) The corresponding spectra in the frequency domain exhibit clear fifth- and third-harmonic generation in addition to the signal of the fundamental frequency.
The frequency-domain spectra of the emitted electric fields (\textbf{c}) from the cavity and (\textbf{d}) from the bare Cd$_3$As$_2$ thin-film samples for various driving electric fields from to 29 to 72~kV/cm.
(\textbf{e}) The intensity of the fifth-harmonic yield versus the driving pulse intensity for the bare and the cavity-engineered Cd$_3$As$_2$ samples.
For the bare Cd$_3$As$_2$, the fifth-harmonic generation at lower drive intensities can hardly be resolved within the experimental uncertainties.
In contrast, the fifth-harmonic generation in the heterostructure is enhanced by more than two orders of magnitude, whose fluence dependence follows $I_{5f} \propto I^{1.9}_f$ towards the highest field (dotted line). This deviates far from $I_{5f} \propto I^{5}_f$, hence the system is deep in the nonperturbative nonlinear regime.
}
\end{figure*}

Whereas fifth-order harmonic generation in Cd$_3$As$_2$ was observed previously by using a powerful accelerator-based THz source \cite{Kovalev2020}, here we are able to resolve the fifth-order harmonic generation based on a table-top laser-based THz source.
Figure~\ref{fig:FHG}a and \ref{fig:FHG}b show the time-domain signals and the frequency-domain spectra of the emission in the bare and in the microcavity samples driven by a highest available THz peak field of 72~kV/cm.
While the fifth-harmonic generation of the bare Cd$_3$As$_2$ thin film is merely resolved with a signal-to-noise ratio of three, the microcavities enhance the fifth-harmonic intensity by more than two orders of magnitude.
A third-harmonic signal can also be detected even after a $5f$-bandpass filter for the microcavity sample, whereas for the reference Cd$_3$As$_2$ thin film it is too weak to be resolved above the noise floor (see Figure~\ref{fig:FHG}b).

We further investigate the cavity-enhanced fifth-harmonic generation by varying the driving field strength. The results are presented in Figure~\ref{fig:FHG}c and~\ref{fig:FHG}d. 
For the thin film sample without microcavities the fifth-harmonic intensity is very weak at lower driving fields, hence cannot be followed below a peak driving field of 54~kV/cm.
In contrast, the microcavity enhanced fifth-harmonic generation can be well resolved for a driving field down to 29~kV/cm.
The dependencies of the fifth-harmonic yields on the driving pulse intensity are summarized in Figure~\ref{fig:FHG}e.
The microcavity enhanced fifth-harmonic generation is found to be deeply into the nonperturbative regimes. The observed power-law dependencies vary from $I_{5f} \propto I^{2.5}_f$ at lower fluences to $I_{5f} \propto I^{1.9}_f$ at higher fluences (see the power-law fits indicated by the dotted lines in Figure~\ref{fig:FHG}e), in contrast to the perturbative $I^{5}_f$ law (dashed line).
Although the resolved third-harmonic generation indicates a nonperturbative response of the Dirac fermions in this material, we can still extrapolate the perturbative $I^{5}_f$ dependence of the fifth-harmonic generation to the lowest fluence for comparison (see dashed line in Figure~\ref{fig:FHG}e), which shows a cavity-induced enhancement also for the fifth-harmonic generation by orders of magnitude. With the table-top laser-based THz source, however, we have not been able to reach this regime due to the limited signal-to-noise. An accelerator-based source with higher repetition rate would be more suitable \cite{Kovalev2020}.

In conclusion, we have studied strong THz field-driven nonthermal dynamics of Dirac fermions in the three-dimensional Dirac semimetal Cd$_3$As$_2$, by investigating their THz third- and fifth-order harmonic generation.
By utilizing designed microcavities, we have been able to modify interactions of THz electrodynamic waves with the Dirac semimetal, thereby engineering the high-order nonlinear responses of the Dirac fermions. 
In particular, the substantially increased local THz field by the microcavity metasurface allows us to reach the near-saturation nonperterbative nonlinear regime, which confirms our predictions for the field-driven nonthermal dynamics of Dirac fermions by using the Boltzmann transport theory.
The metasurface microcavities allow us to enhance the high-order nonlinear THz responses by several ten times up to more than three orders of magnitude.
Since materials with strong and sensitive THz responses are crucial for the development of sixth-generation communication technology, the observed room-temperature and highly efficient THz high-harmonic generation in the engineered sample is particularly relevant for optoelectronic applications \cite{Deinert2021,Tielrooij2022,DiGaspare2024,
deabajo2025roadmapphotonics2dmaterials}.
Our study demonstrates the fruitful possibilities of engineering nonlinear optical responses of three-dimensional Dirac fermions and beyond in quantum materials \cite{SentefPRR2020,Huebener2021,JuraschekPRR2021}.
\\

Supporting information is available online: https://doi.org/10.1021/acs.nanolett.6c03104

\section*{Acknowledgements}

We acknowledge support by the European Research Council (ERC) under the Horizon 2020 research and innovation programme, Grant Agreement No. 950560 (DynaQuanta), and by the European Union through the FET Open project EXTREME IR (944735).
S.D. was supported by the Walter Benjamin Programme of the German Research Foundation (DFG) via the Project Number 557566396.
F.X. was supported by the National Natural Science Foundation of China (52225207 and 52350001), the Shanghai Pilot Program for Basic Research - FuDan University 21TQ1400100 (21TQ006) and the Shanghai Municipal Science and Technology Major Project (Grant No.2019SHZDZX01). 

S.D. and L.S. contributed equally to this work.

\printbibliography

\clearpage

\end{document}